\documentclass[12pt]{iopart}
  \expandafter\let\csname equation*\endcsname\relax
  \expandafter\let\csname endequation*\endcsname\relax 
\usepackage[utf8x]{inputenc}
\usepackage{amsmath,amsfonts,mathrsfs,amssymb,dsfont,braket,bbold,graphicx,float}
\usepackage{lineno,color}
  \makeatletter
  \let\start@align@nopar\start@align
  \let\start@gather@nopar\start@gather
  \let\start@multline@nopar\start@multline
  \long\def\start@align{\par\start@align@nopar}
  \long\def\start@gather{\par\start@gather@nopar}
  \long\def\start@multline{\par\start@multline@nopar}
  \makeatother
\definecolor{darkblue}{rgb}{0,0,.6}
\usepackage{hyperref}
\hypersetup{colorlinks=true,linkcolor=darkblue,citecolor=darkblue,urlcolor=darkblue}

\begin{document}

% LINE NUMBERING
%\linenumbers

%%%%%%%%%%%%%%%%%%%%%%%%%%%%%%%%%%%%%%%%%%%%%%%%%%%%%%%%%%%%%%%%%%%%%%%%%%%%%%%%%%%%%%%%%%%%%%%%%%%%%%%%%%%%%%%%%%%%%%%%%%%%%%%%%%%
%	TITLE
%%%%%%%%%%%%%%%%%%%%%%%%%%%%%%%%%%%%%%%%%%%%%%%%%%%%%%%%%%%%%%%%%%%%%%%%%%%%%%%%%%%%%%%%%%%%%%%%%%%%%%%%%%%%%%%%%%%%%%%%%%%%%%%%%%%

\title[Entropic uncertainty bound for open pointer-based measurements]{Entropic uncertainty bound for open pointer-based simultaneous measurements of conjugate observables}
\author{Raoul Heese and Matthias Freyberger}
\address{Institut f{\"u}r Quantenphysik and Center for Integrated Quantum Science and Technology (IQ\textsuperscript{ST}), Universit{\"a}t Ulm, D-89069 Ulm, Germany}
\ead{raoul.heese@uni-ulm.de}
\begin{abstract}We discuss the information entropy for a general open pointer-based simultaneous measurement and show how it is bound from below. This entropic uncertainty bound is a direct consequence of the structure of the entropy and can be obtained from the formal solution of the measurement dynamics. Furthermore, the structural properties of the entropy allow us to give an intuitive interpretation of the noisy influence of the pointers and the environmental heat bath on the measurement results. \par\noindent{\it Keywords\/}: simultaneous pointer-based measurement, noisy measurement, conjugate observables, entropy, uncertainty relation, quantum mechanics\end{abstract}
\maketitle

%%%%%%%%%%%%%%%%%%%%%%%%%%%%%%%%%%%%%%%%%%%%%%%%%%%%%%%%%%%%%%%%%%%%%%%%%%%%%%%%%%%%%%%%%%%%%%%%%%%%%%%%%%%%%%%%%%%%%%%%%%%%%%%%%%%
%	CONTENT
%%%%%%%%%%%%%%%%%%%%%%%%%%%%%%%%%%%%%%%%%%%%%%%%%%%%%%%%%%%%%%%%%%%%%%%%%%%%%%%%%%%%%%%%%%%%%%%%%%%%%%%%%%%%%%%%%%%%%%%%%%%%%%%%%%%

\section{Introduction} \label{sec:Introduction}
The concept of pointer-based simultaneous measurements of conjugate observables is an indirect measurement model, which allows to dynamically describe the properties of a simultaneous quantum mechanical measurement process. Additional to the system to be measured (hereafter just called \emph{system}), the model introduces two additional systems called \emph{pointers}, which are coupled to the system and act as commuting meters from which the initial system observables can be read out after a certain interaction time. In this sense, the pointers represent the measurement devices used to simultaneously determine the system observables. Pointer-based simultaneous measurements date back to Arthurs and Kelly \cite{arthurs1965} and are based on von Neumann's idea of indirect observation \cite{vonneumann1932}.\par
\begin{figure}[H]
  \centering \includegraphics{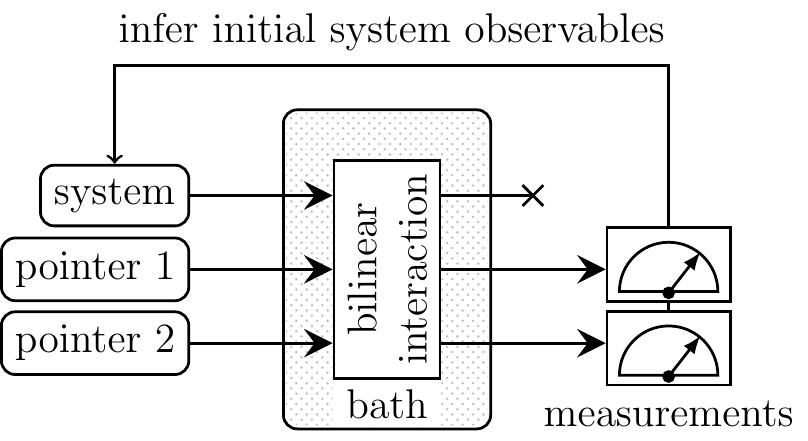}
  \caption{Principles of an open pointer-based simultaneous measurement of two conjugate observables, e.\,g., the simultaneous measurement of position and momentum. The measurement apparatus consists of two quantum mechanical systems, called pointers, which are bilinearly coupled to the quantum mechanical system to be measured. Additionally, an environmental heat bath in the sense of the Caldeira-Leggett model can disturb both the system and the pointers. After the interaction process, one observable of each pointer is directly measured (e.\,g., the position of each pointer) while the system itself is not subject to any direct measurement. However, from these measurement results, information about the initial system observables can then be inferred. In other words, the final pointer observables act as commuting meters from which the initial non-commuting system observables can be simultaneously read out. The price to be paid for this simultaneity comes in form of fundamental noise terms, which affect the inferred values. The corresponding uncertainties can be described by information entropies, which are bound from below.}
  \label{fig:model}
\end{figure}

In principle, any pair of conjugate observables like position and momentum or quadratures of the electromagnetic field \cite{schleich2001}, whose commutator is well-defined and proportional to the identity operator, can straightforwardly be measured within the scope of pointer-based simultaneous measurements. We limit ourselves to the measurement of position and momentum in the following. An \emph{open} pointer-based simultaneous measurement \cite{heese2014} also takes environmental effects into consideration by utilizing an environmental heat bath in the sense of the Caldeira-Leggett model \cite{caldeira1981,caldeira1983a,caldeira1983ae,caldeira1983b}, which leads to a quantum Brownian motion \cite{chou2008,fleming2011a,fleming2011b,martinez2013} of the system and the pointers, whereas a \emph{closed} pointer-based simultaneous measurement \cite{arthurs1965,wodkiewicz1984,stenholm1992,buzek1995,appleby1998a,appleby1998b,appleby1998c,busch2007,busshardt2010,busshardt2011,heese2013} does not involve any environmental effects. A schematic open pointer-based simultaneous measurement procedure is shown in Fig.~\ref{fig:model}.\par
In this contribution we calculate the information entropy of an open pointer-based simultaneous measurement and discuss its properties as a measurement uncertainty. In particular, we make use of recent results \cite{heese2013,heese2014}, which we extend and generalize. In Sec.~\ref{sec:generalopenpointer-basedsimultaneousmeasurements}, we present the formal dynamics of open pointer-based simultaneous measurements and then use these results to discuss the entropic uncertainty in Sec.~\ref{sec:entropy}. In the end, we arrive at a generic lower bound of this entropic uncertainty. Note that we solely use rescaled dimensionless variables \cite{heese2014} so that $\hbar = 1$.

%%%%%%%%%%%%%%%%%%%%%%%%%%%%%%%%%%%%%%%%%%%%%%%%%%%%%%%%%%%%%%%%%%%%%%%%%%%%%%%%%%%%%%%%%%%%%%%%%%%%%%%%%%%%%%%%%%%%%%%%%%%%%%%%%%%
%%%%%%%%%%%%%%%%%%%%%%%%%%%%%%%%%%%%%%%%%%%%%%%%%%%%%%%%%%%%%%%%%%%%%%%%%%%%%%%%%%%%%%%%%%%%%%%%%%%%%%%%%%%%%%%%%%%%%%%%%%%%%%%%%%%

\section{Open pointer-based simultaneous measurements} \label{sec:generalopenpointer-basedsimultaneousmeasurements}
As indicated in the introduction, our model of open pointer-based simultaneous measurements consists of a system particle to be measured with mass $M_{\mathrm{S}}$, position observable $\hat{X}_{\mathrm{S}}$ and momentum observable $\hat{P}_{\mathrm{S}}$, which is coupled bilinearly to two pointer particles with masses $M_1$ and $M_2$, position observables $\hat{X}_1$ and $\hat{X}_2$, and momentum observables $\hat{P}_1$ and $\hat{P}_2$, respectively. Both the system and the pointers are bilinearly coupled to an environmental heat bath, which consists of a collection of $N$ harmonic oscillators with masses $m_1,\dots,m_N$, position observables $\hat{q}_1,\dots,\hat{q}_N$, and momentum observables $\hat{k}_1,\dots,\hat{k}_N$. In this section, we first present the general Hamiltonian for this model and then briefly discuss the resulting dynamics.

\subsection{Hamiltonian} \label{sec:generalopenpointer-basedsimultaneousmeasurements:hamiltonian}
The general Hamiltonian for our model reads
\begin{align} \label{eq:H}
  \hat{\mathscr{H}}(t) \equiv \hat{H}_{\mathrm{free}} + \hat{H}_{\mathrm{int}}(t) + \hat{H}_{\mathrm{bath}}(t)
\end{align}
and therefore consists of three parts. First, the free evolution Hamiltonian
\begin{align} \label{eq:H:free}
  \hat{H}_{\mathrm{free}} \equiv \frac{\hat{P}_{\mathrm{S}}^2}{2 M_{\mathrm{S}}} + \frac{\hat{P}_1^2}{2 M_1} + \frac{\hat{P}_2^2}{2 M_2},
\end{align}
which simply describes the dynamics of the undisturbed system and pointers. Second, the interaction Hamiltonian
\begin{align} \label{eq:H:int}
  \hat{H}_{\mathrm{int}}(t) \equiv C_{\mathrm{S}}(t) \hat{X}_{\mathrm{S}}^2 + C_1(t) \hat{X}_1^2 + C_2(t) \hat{X}_2^2 + ( \hat{X}_{\mathrm{S}}, \hat{P}_{\mathrm{S}} ) \mathbf{C}(t) ( \hat{X}_1, \hat{X}_2, \hat{P}_1, \hat{P}_2 )^T,
\end{align}
which describes possible quadratic potentials with the coupling strengths $C_{\mathrm{S}}(t)$, $C_1(t)$, and $C_2(t)$, respectively, as well as bilinear interactions between the system observables and the pointer observables via the $2\times4$ coupling matrix $\mathbf{C}(t)$. These interactions are necessary for an information transfer between system and pointers and are therefore a prerequisite of pointer-based simultaneous measurements. The existence of quadratic potentials is on the other hand not essential, but may be reasonable from a physical point of view when regarding confined particles. One possible interaction Hamiltonian would be the interaction Hamiltonian of the classic Arthurs and Kelly model \cite{arthurs1965}, which can be written as $\hat{H}_{\mathrm{int}} = \kappa ( \hat{X}_{\mathrm{S}} \hat{P}_1 + \hat{P}_{\mathrm{S}} \hat{P}_2 )$ with an arbitrary coupling strength $\kappa \neq 0$.\par
Lastly, Eq.~{(\ref{eq:H})} contains the bath Hamiltonian \cite{caldeira1981,caldeira1983a,caldeira1983ae,caldeira1983b}
\begin{align} \label{eq:H:bath}
  \hat{H}_{\mathrm{bath}}(t) \equiv \frac{1}{2} \hat{\mathbf{k}}^{T} \mathbf{m}^{-1} \hat{\mathbf{k}}+ \frac{1}{2} \hat{\mathbf{q}}^{T} \mathbf{c} \hat{\mathbf{q}} + \hat{\mathbf{q}}^{T} \mathbf{g}(t) ( \hat{X}_{\mathrm{S}}, \hat{X}_1, \hat{X}_2 )^T,
\end{align}
which describes the independent dynamics of the bath particles with the $N \times N$ diagonal mass matrix $\mathbf{m}$ containing $m_1,\dots,m_N$, and the $N \times N$ symmetric and positive definite bath-internal coupling matrix $\mathbf{c}$; as well as the coupling of system and pointer positions to the bath positions with the coupling strength
\begin{align} \label{eq:g}
  \mathbf{g}(t) \equiv g(t) \mathbf{g},
\end{align}
which consists of the time-dependent scalar $g(t)$ and the time-independent $N\times 3$ matrix $\mathbf{g}$. To simplify our notation, we make use of the vectorial bath positions $\hat{\mathbf{q}}^{T} \equiv (\hat{q}_1, \dots, \hat{q}_N)$ and the vectorial bath momenta $\hat{\mathbf{k}}^{T} \equiv (\hat{k}_1, \dots, \hat{k}_N)$.\par
In particular, since all of the following calculations only rely on the bilinear structure of the Hamiltonian, Eq.~{(\ref{eq:H})}, we do not further specify the coupling strengths and therefore consider quite general measurement configurations. Note that the time-dependencies of the coupling strengths in Eqs.~{(\ref{eq:H:int})} and {(\ref{eq:H:bath})} allow us to design specific coupling pulses for system-pointer interactions \cite{busshardt2010} or switch-on functions for the bath \cite{fleming2011a,fleming2011c}.

\subsection{Dynamics}
To determine the complete system and pointer dynamics, it is necessary to solve the coupled Heisenberg equations of motion
\begin{align} \label{eq:Heisenberg}
  & \phantom{=.} \frac{\partial}{\partial t} \left( \hat{X}_{\mathrm{S}}(t), \hat{X}_1(t), \hat{X}_2(t), \hat{P}_{\mathrm{S}}(t), \hat{P}_1(t), \hat{P}_2(t) \right) \nonumber \\
  & = i \left[ \hat{\mathscr{H}}(t), \left( \hat{X}_{\mathrm{S}}(t), \hat{X}_1(t), \hat{X}_2(t), \hat{P}_{\mathrm{S}}(t), \hat{P}_1(t), \hat{P}_2(t) \right) \right],
\end{align}
which take on the form of a Volterra integro-differential equation \cite{burton2005} after explicitly solving the dynamics of the bath particles \cite{ford1988,fleming2011a,fleming2011b}. The general solution of Eq.~{(\ref{eq:Heisenberg})} can formally be expressed by means of a resolvent \cite{grossmann1968,becker2006} and leads to system and pointer positions and momenta, which are linearly propagated from their initial values (i.\,e., the homogeneous solution) and are affected by an additive noise term (i.\,e., the inhomogeneous solution). Explicit analytical solutions, which are naturally of the same structure as the formal solution, can only be found for specific choices of the Hamiltonian, Eq.~{(\ref{eq:H})}. Otherwise, numerical approaches \cite{chen1998,baker2000} might be necessary. Interestingly, since we strive after a general discussion of the dynamics, the ensured existence of a solution and its formal structure is sufficient for all further considerations. We assume in this context that possible unphysical artifacts of the modeling \cite{weiss1999,fleming2011a}, e.\,g., renormalization problems, have been treated adequately \cite{heese2014}. Without loss of generality, we choose $t = 0$ as the initial time.\par
For the description of a pointer-based simultaneous measurement, we do in fact not need knowledge about the complete system and pointer dynamics. As outlined in Fig.~\ref{fig:model}, the measurement process provides that the two pointers are being measured after an interaction time $t$ in such a way, that we either read out the position or the momentum of each pointer. Consequently, we have knowledge about either $\hat{X}_1(t)$ or $\hat{P}_1(t)$ of the first pointer, and, additionally, about either $\hat{X}_2(t)$ or $\hat{P}_2(t)$ of the second pointer, which is a total of four different possible measurement combinations. To summarize these measurement combinations, we define the measurement vector $\hat{\mathbf{w}}(t)$, which consists of the two pointer observables chosen to be read out, e.\,g., $\hat{\mathbf{w}}(t) = (\hat{X}_1(t), \hat{X}_2(t))^T$ when measuring both pointer positions. As also outlined in Fig.~\ref{fig:model}, the information gained from measuring the observables in the measurement vector allows us to infer the initial system observables $\hat{X}_{\mathrm{S}}(0)$ and $\hat{P}_{\mathrm{S}}(0)$. For this reason, the connection between the measured observables after a certain interaction time $t$ and the initial system observables builds the framework for a description of pointer-based simultaneous measurements.\par
With this purpose in mind, we can extract
\begin{align} \label{eq:inferredsystem}
  \begin{pmatrix} \hat{\mathcal{X}}(t) \\ \hat{\mathcal{P}}(t) \end{pmatrix} = \begin{pmatrix} \hat{X}_{\mathrm{S}}(0) \\ \hat{P}_{\mathrm{S}}(0) \end{pmatrix} + \mathbf{B}(t) \hat{\mathbf{J}} + (\mathbf{\Lambda} \star \hat{\boldsymbol{\xi}})(t)
\end{align}
from any (formal) solution of the Heisenberg equations of motion, Eq.~{(\ref{eq:Heisenberg})}. Here we have introduced the so-called generalized inferred observables
\begin{align} \label{eq:inferredobservables}
\begin{pmatrix} \hat{\mathcal{X}}(t) \\ \hat{\mathcal{P}}(t) \end{pmatrix} \equiv \mathbf{A}(t) \hat{\mathbf{w}}(t),
\end{align}
which are given by a rescaled measurement vector $\hat{\mathbf{w}}(t)$ and can on the other hand be understood as the effectively measured observables from which the system observables can be directly read out. Since the two measured pointer observables in $\hat{\mathbf{w}}(t)$ initially commute and are subject to a unitary evolution, one has
\begin{align} \label{eq:XPcommute}
  [\hat{\mathcal{X}}(t),\hat{\mathcal{P}}(t)] = 0,
\end{align}
which means that the inferred observables can be determined simultaneously.\par
Furthermore, Eqs.~{(\ref{eq:inferredsystem})} and {(\ref{eq:inferredobservables})} contain the coefficient matrices $\mathbf{A}(t)$, $\mathbf{B}(t)$, and $\mathbf{\Lambda}(t,s)$ with $0 \leq s \leq t$, for which we do not need to give an explicit expression. In general, they can be straightforwardly calculated from the resolvent of the formal solution of the complete system and pointer dynamics. An exemplary calculation can be found in Ref.~\cite{heese2014}. We also make use of the initial value vector
\begin{align} \label{eq:J}
  \hat{\mathbf{J}} \equiv ( \hat{X}_1(0), \hat{X}_2(0), \hat{P}_1(0), \hat{P}_2(0) )^{T},
\end{align}
and the stochastic force \cite{weiss1999}
\begin{align} \label{eq:stochasticforce}
  \hat{\boldsymbol{\xi}}(t) \equiv -\mathbf{g}^T(t) \left( \mathbf{m}^{-\frac{1}{2}} \cos( \boldsymbol{\omega} t) \mathbf{m}^{\frac{1}{2}} \hat{\mathbf{q}} + \mathbf{m}^{-\frac{1}{2}} \sin( \boldsymbol{\omega} t) \boldsymbol{\omega}^{-1} \mathbf{m}^{-\frac{1}{2}} \hat{\mathbf{k}} \right)
\end{align}
with the symmetric bath frequency matrix \cite{fleming2011b}
\begin{align} \label{eq:omega}
  \boldsymbol{\omega} \equiv \sqrt{ \mathbf{m}^{-\frac{1}{2}} \mathbf{c} \mathbf{m}^{-\frac{1}{2}} }.
\end{align}
In particular, the stochastic force results from the homogeneous bath dynamics and describes the noisy influence of the bath on the measurement results, Eq.~{(\ref{eq:H:bath})}. The symbol $\star$ in Eq.~{(\ref{eq:inferredsystem})} represents the integral
\begin{align} \label{eq:star}
  (f \star g)(x) \equiv \int \limits_{0}^{x} \! \mathrm d y f(x,y) g(y)
\end{align}
for two arbitrary functions $f(x,y)$ and $g(x)$. In case of $f(x,y) = f(x-y)$, Eq.~{(\ref{eq:star})} is called a Laplace convolution.\par
The central aspect of the dynamics contained in Eq.~{(\ref{eq:inferredsystem})} can be understood when considering the expectation value
\begin{align} \label{eq:inferredexpecation}
  \Braket{ \begin{pmatrix} \hat{\mathcal{X}}(t) \\ \hat{\mathcal{P}}(t) \end{pmatrix} } = \Braket{ \begin{pmatrix} \hat{X}_{\mathrm{S}}(0) \\ \hat{P}_{\mathrm{S}}(0) \end{pmatrix} } + \mathbf{s}(t)
\end{align}
of Eq.~{(\ref{eq:inferredsystem})}, where the brackets refer to the mean value with respect to the initial state. Most importantly, Eq.~{(\ref{eq:inferredexpecation})} shows that the knowledge about the first moments of the inferred observables, Eq.~{(\ref{eq:inferredobservables})}, which can be determined by measuring the two pointer observables contained in the measurement vector $\hat{\mathbf{w}}(t)$, allows us to infer the first moments of the initial system observables. The second and third term on the right-hand side of Eq.~{(\ref{eq:inferredsystem})} simply add the shift 
\begin{align} \label{eq:s}
  \mathbf{s}(t) \equiv \mathbf{B}(t) \braket{ \hat{\mathbf{J}} } + (\mathbf{\Lambda} \star \braket{ \hat{\boldsymbol{\xi}}})(t)
\end{align}
to this relation. Assuming a suitably separable initial state, this shift is determined by the initial state of the pointers and the bath, but independent of the state of the system. In this sense, our model is only of any practical purpose if the initial pointer and bath states are sufficiently well-known to determine $\mathbf{s}(t)$. Generally, the initial pointer states are at our disposal and can therefore be chosen in such a way that the first term on the right-hand side of Eq.~{(\ref{eq:s})} vanishes. It is furthermore reasonable to suppose that an environmental heat bath in thermal equilibrium leads to a vanishing expectation value of the stochastic force $\hat{\boldsymbol{\xi}}(t)$, Eq.~{(\ref{eq:stochasticforce})}. Consequently, presuming a vanishing shift $\mathbf{s}(t)$ seems appropriate for a typical measurement configuration. We will confirm this presumption further below for a specifically chosen initial state. Note that the linearity of Eqs.~{(\ref{eq:inferredsystem})} and {(\ref{eq:inferredobservables})}, which is essential for the inference process, Eq.~{(\ref{eq:inferredexpecation})}, is a direct result of the bilinear structure of the Hamiltonian, Eq.~{(\ref{eq:H})}.\par
We assume here the non-pathological case that the interaction Hamiltonian $\hat{H}_{\mathrm{int}}(t)$, Eq.~{(\ref{eq:H:int})}, and the measurement vector $\hat{\mathbf{w}}(t)$ are chosen in such a way that inferring system observables from the pointers is possible in the first place, which is defined by the existence of the coefficient matrix $\mathbf{A}(t)$, Eq.~{(\ref{eq:inferredobservables})}. In other words, we require a sufficient information transfer from the system to the pointers. For example, for the classic Arthurs and Kelly model \cite{arthurs1965} with the interaction Hamiltonian mentioned in Sec.~\ref{sec:generalopenpointer-basedsimultaneousmeasurements:hamiltonian} and no environmental heat bath, measuring both pointer positions leads to an existing matrix $\mathbf{A}(t)$ for $t>0$, but measuring both pointer momenta does not \cite{busshardt2010}.\par
Seeing now the role played by the inferred observables, we can quantify the uncertainty of a simultaneous pointer-based measurement with the help of the so-called noise operators \cite{arthurs1988}
\begin{align} \label{eq:noiseoperators}
  \begin{pmatrix} \hat{N}_{\mathcal{X}}(t) \\ \hat{N}_{\mathcal{P}}(t) \end{pmatrix} \equiv \begin{pmatrix} \hat{\mathcal{X}}(t) - \hat{X}_{\mathrm{S}}(0) \\ \hat{\mathcal{P}}(t) - \hat{P}_{\mathrm{S}}(0) \end{pmatrix} = \mathbf{B}(t) \hat{\mathbf{J}} + (\mathbf{\Lambda} \star \hat{\boldsymbol{\xi}})(t),
\end{align}
which are defined as the difference between the inferred observables $\hat{\mathcal{X}}(t)$ and $\hat{\mathcal{P}}(t)$, Eq.~{(\ref{eq:inferredobservables})}, and the respective initial system observables $\hat{X}_{\mathrm{S}}(0)$ and $\hat{P}_{\mathrm{S}}(0)$. The second equal sign in Eq.~{(\ref{eq:noiseoperators})} directly follows from Eq.~{(\ref{eq:inferredsystem})}. The expectation values
\begin{align} \label{eq:noiseoperators:exp}
  \Braket{ \begin{pmatrix} \hat{N}_{\mathcal{X}}(t) \\ \hat{N}_{\mathcal{P}}(t) \end{pmatrix} } = \mathbf{s}(t)
\end{align}
of the noise operators correspond to the shift $\mathbf{s}(t)$, Eq.~{(\ref{eq:s})}. As we have mentioned above, this shift can typically be presumed to vanish. Particularly interesting for our considerations is the symmetrized covariance matrix
\begin{align} \label{eq:noiseoperators:cov}
  \begin{pmatrix} \Braket{ \hat{N}_{\mathcal{X}}^2(t) } - \Braket{ \hat{N}_{\mathcal{X}}(t) }^2 & \Braket{\hat{N}_{\mathcal{X}\mathcal{P}}(t)} - \Braket{ \hat{N}_{\mathcal{X}}(t) }\Braket{ \hat{N}_{\mathcal{P}}(t) } \\ \Braket{\hat{N}_{\mathcal{X}\mathcal{P}}(t)} - \Braket{ \hat{N}_{\mathcal{X}}(t) }\Braket{ \hat{N}_{\mathcal{P}}(t) } & \Braket{ \hat{N}_{\mathcal{P}}^2(t) } - \Braket{ \hat{N}_{\mathcal{P}}(t) }^2 \end{pmatrix} \equiv \begin{pmatrix} \delta_{\mathcal{X}}^2(t) & \delta_{\mathcal{X}\mathcal{P}}(t) \\ \delta_{\mathcal{X}\mathcal{P}}(t) & \delta_{\mathcal{P}}^2(t) \end{pmatrix}
\end{align}
of the noise operators, where
\begin{align} \label{eq:noiseoperators:NXP}
  \hat{N}_{\mathcal{X}\mathcal{P}}(t) \equiv \frac{1}{2} \left( \hat{N}_{\mathcal{X}}(t)\hat{N}_{\mathcal{P}}(t) + \hat{N}_{\mathcal{P}}(t)\hat{N}_{\mathcal{X}}(t) \right)
\end{align}
stands for the symmetrized noise operator. To simplify our notation, we have also introduced the noise terms $\delta_{\mathcal{X}}^2(t)$ and $\delta_{\mathcal{P}}^2(t)$, which represent the diagonal elements of the covariance matrix, and the correlation term $\delta_{\mathcal{X}\mathcal{P}}(t)$, which represents the off-diagonal elements. Since a precise knowledge of $\mathbf{s}(t)$, Eqs.~{(\ref{eq:s})} and {(\ref{eq:noiseoperators:exp})}, is necessary for the success of the inference process, Eq.~{(\ref{eq:inferredexpecation})}, the associated variances given by the noise terms $\delta_{\mathcal{X}}^2(t)$ and $\delta_{\mathcal{P}}^2(t)$ can be considered as an intuitive measure for the respective measurement uncertainty. Indeed, in the scope of variance-based uncertainty relations, the noise terms play an important role; see, e.\,g., Ref.~\cite{ozawa2005} and references therein. Note that the noise terms are also referred to as ``errors of retrodiction" \cite{appleby1998a,appleby1998b,appleby1998c}. We will revisit them further below, where we will see that they arise naturally in our entropic description of the measurement uncertainty.\par
The formal dynamics of the inferred observables $\hat{\mathcal{X}}(t)$ and $\hat{\mathcal{P}}(t)$, Eq.~{(\ref{eq:inferredsystem})}, build the framework for all of the following considerations. Most importantly, the intimate connection between $\hat{\mathcal{X}}(t)$ and $\hat{\mathcal{P}}(t)$, which result from the measured pointer observables in $\hat{\mathbf{w}}(t)$, Eq.~{(\ref{eq:inferredobservables})}, and the initial system observables $\hat{X}_{\mathrm{S}}(0)$ and $\hat{P}_{\mathrm{S}}(0)$ can clearly be seen from Eq.~{(\ref{eq:inferredsystem})}. It is this connection which lies at the heart of the pointer-based measurement scheme: information about the non-commuting system observables can be gathered from knowledge about the commuting inferred observables. At this point it seems natural to ask: What is the uncertainty of such an indirect measurement process? In the next section, we answer this question with the help of information entropy.

%%%%%%%%%%%%%%%%%%%%%%%%%%%%%%%%%%%%%%%%%%%%%%%%%%%%%%%%%%%%%%%%%%%%%%%%%%%%%%%%%%%%%%%%%%%%%%%%%%%%%%%%%%%%%%%%%%%%%%%%%%%%%%%%%%%
%%%%%%%%%%%%%%%%%%%%%%%%%%%%%%%%%%%%%%%%%%%%%%%%%%%%%%%%%%%%%%%%%%%%%%%%%%%%%%%%%%%%%%%%%%%%%%%%%%%%%%%%%%%%%%%%%%%%%%%%%%%%%%%%%%%

\section{Entropy} \label{sec:entropy}
The uncertainty principle \cite{heisenberg1927,folland1997,busch2007} is of central importance for quantum mechanical measurements. It manifests itself in the form of uncertainty relations \cite{busch2014}, usually written in terms of variances \cite{busch2013a,busch2013b} or information entropies \cite{buscemi2014,coles2014}. The concept of information entropies goes back to Ref.~\cite{shannon1948}, whereas its general usage in the scope of uncertainty relations has been pioneered by Ref.~\cite{hirschman1957,beckner1975,bialynicki1975}. In the context of closed pointer-based simultaneous measurements \cite{arthurs1965,stenholm1992}, the uncertainty principle leads to well-known variance-based uncertainty relations \cite{appleby1998a,appleby1998b,appleby1998c}, which have extensively been discussed, e.\,g., in the scope of phase-space measurements \cite{wodkiewicz1984} or energy and timing considerations \cite{busshardt2010,busshardt2011}. A respective information entropic form has been derived in Ref.~\cite{buzek1995} and improved in Ref.~\cite{heese2013}. Variances can also be used to describe an uncertainty relation for open pointer-based simultaneous measurements \cite{heese2014}. However, in comparison with information entropies, variances suffer from two major drawbacks \cite{hilgevoord1990,buzek1995,bialynicki2011}: First, they can become divergent for specific probability distributions and second, they may not reflect what one would intuitively consider as the ``width" of a probability distribution. On the other hand, a disadvantage of information entropies is that they are usually much more difficult to calculate than variances. This, however, is only a technical limitation. Therefore, we concentrate on information entropic uncertainties in the following.\par
In this section, we first introduce the so-called collective entropy as a total measure of uncertainty in the context of open pointer-based simultaneous measurements. Choosing a separable initial state then allows us to calculate the marginal probability distributions this collective entropy is based on. As a result, we can discuss the structural properties of the collective entropy, including an extension of a previously known lower bound \cite{heese2013}.

\subsection{Collective entropy}
First concepts of information entropies as an uncertainty measure for pointer-based simultaneous measurements can be found in Ref.~\cite{buzek1995}, which serves as a foundation for the present section. Since information entropies are based on probability distributions, we first need to define suitable probability distributions before we can deal with the actual entropies.\par
The probability of measuring an inferred position $\mathcal{X}$ and an inferred momentum $\mathcal{P}$ is given by the joint probability distribution \cite{wodkiewicz1984,wodkiewicz1986}
\begin{align} \label{eq:prXP}
  \mathrm{pr}(\mathcal{X},\mathcal{P};t) \equiv \Braket{ \delta(\mathcal{X}-\hat{\mathcal{X}}(t)) \delta(\mathcal{P}-\hat{\mathcal{P}}(t)) }
\end{align}
with the Dirac delta distributions $\delta(\mathcal{X}-\hat{\mathcal{X}}(t))$ and $\delta(\mathcal{P}-\hat{\mathcal{P}}(t))$, which contain the inferred observables $\hat{\mathcal{X}}(t)$ and $\hat{\mathcal{P}}(t)$, Eq.~{(\ref{eq:inferredobservables})}, respectively. Likewise, the probability of measuring either the inferred position $\mathcal{X}$ or the inferred momentum $\mathcal{P}$ is given by the marginal probability distribution of inferred position
\begin{subequations} \label{eq:Pmarginal}
  \begin{align}
    \mathrm{pr}_{\mathcal{X}}(\mathcal{X};t) \equiv \int \limits_{- \infty}^{+ \infty} \! \mathrm d \mathcal{P}\ \mathrm{pr}(\mathcal{X},\mathcal{P};t)
  \end{align}  
and inferred momentum
  \begin{align}
    \mathrm{pr}_{\mathcal{P}}(\mathcal{P};t) \equiv \int \limits_{- \infty}^{+ \infty} \! \mathrm d \mathcal{X}\ \mathrm{pr}(\mathcal{X},\mathcal{P};t),
  \end{align}
\end{subequations}
respectively.\par
With the help of the marginal probability distributions, Eq.~{(\ref{eq:Pmarginal})}, one can define the collective entropy \cite{heese2013}
\begin{align} \label{eq:S}
  S(t) \equiv S_{\mathcal{X}}(t) + S_{\mathcal{P}}(t),
\end{align} 
which describes the total uncertainty of a simultaneous pointer-based measurement process. It consists of the sum of the marginal entropy of inferred position
\begin{subequations} \label{eq:Smarginal}
\begin{align} \label{eq:Smarginal:X}
  S_{\mathcal{X}}(t) \equiv - \int \limits_{- \infty}^{+ \infty} \! \mathrm d \mathcal{X}\ \mathrm{pr}_{\mathcal{X}}(\mathcal{X};t) \ln \mathrm{pr}_{\mathcal{X}}(\mathcal{X};t)
\end{align}
and the marginal entropy of inferred momentum
\begin{align} \label{eq:Smarginal:P}
  S_{\mathcal{P}}(t) \equiv - \int \limits_{- \infty}^{+ \infty} \! \mathrm d \mathcal{P}\ \mathrm{pr}_{\mathcal{P}}(\mathcal{P};t) \ln \mathrm{pr}_{\mathcal{P}}(\mathcal{P};t).
\end{align}
\end{subequations}
So far, our considerations have been completely general.

\subsection{Separable initial state} \label{sec:entropy:separable initial state}
A more detailed calculation of the collective entropy, Eq.~{(\ref{eq:S})}, is only possible if we choose a more specific initial state $\hat{\varrho}(0)$ for the measurement configuration. First of all, it is a common approach to assume that the bath is initially in thermal equilibrium and separable from the system and the pointers \footnote{For baths which are not initially separable from the system and the pointers, respectively, our method of calculating the entropy in \ref{sec:appendix:marginal probability distributions} may fail. However, by choosing an appropriate time-dependency of the Hamiltonian, Eq.~{(\ref{eq:H})}, a smooth switch-on of the bath can be introduced to realize a more physically reasonable model; see, e.\,g., Ref.~\cite{fleming2011a,fleming2011c} and references therein.}. Furthermore, it seems natural to choose initially localized and separable pointer states. These localized pointer states are then propagated by the Hamiltonian, Eq.~{(\ref{eq:H})}, in such a way, Eq.~{(\ref{eq:inferredobservables})}, that their new location can be read out at end of the interaction process to infer the system observables, Eq.~{(\ref{eq:inferredsystem})}. A straightforward realization for localized and separable pointer states are squeezed vacuum states \cite{barnett1997,schleich2001}, which have already been used in the classic Arthurs and Kelly model \cite{arthurs1965}. Our system to be measured should, on the other hand, not be subject to any assumptions, so we describe it by a general density matrix.\par
Thus, for our model of open pointer-based simultaneous measurements, we choose a separable initial state
\begin{align} \label{eq:InitialState}
  \hat{\varrho}(0) \equiv \hat{\varrho}_{\mathrm{S}}(0) \otimes \ket{\sigma_1} \bra{\sigma_1} \otimes \ket{\sigma_2} \bra{\sigma_2} \otimes \hat{\varrho}_{\mathrm{B}}(0).
\end{align}
It consists of the initial state $\hat{\varrho}_{\mathrm{S}}(0)$ of the system to be measured, the squeezed vacuum states $\ket{\sigma_1}$ and $\ket{\sigma_2}$ of the pointers, which can be written as \cite{schleich2001}
\begin{align} \label{eq:initialpointerstate}
  \braket{x | \sigma_k} \equiv \left( \frac{1}{2 \pi \sigma_k^2} \right)^{\frac{1}{4}} \exp \left[ - \frac{x^2}{4 \sigma_k^2} \right]
\end{align}
in position space with variances $\sigma_k^2$ for $k \in \{1,2\}$, and the thermal state of the bath
\begin{align} \label{eq:initialbathstate}
  \hat{\varrho}_{\mathrm{B}}(0) \equiv \frac{1}{Z} \exp \left[ - \beta \left\{ \frac{1}{2} \hat{\mathbf{k}}^{T} \mathbf{m}^{-1} \hat{\mathbf{k}}+ \frac{1}{2} \hat{\mathbf{q}}^{T} \mathbf{c} \hat{\mathbf{q}} \right\} \right]
\end{align}
with the thermal energy $\beta^{-1}$ and the normalizing partition function $Z$.\par
The choice of a specific initial state of the bath allows us to determine the statistical properties of the stochastic force $\hat{\boldsymbol{\xi}}(t)$, Eq.~{(\ref{eq:stochasticforce})}. In particular, one has
\begin{align}
 \braket{\hat{\boldsymbol{\xi}}(t)} = 0
\end{align}
and
\begin{align}
  \frac{1}{2} \braket{ \hat{\boldsymbol{\xi}}(t_1) \hat{\boldsymbol{\xi}}{}^{T}(t_2) + \hat{\boldsymbol{\xi}}(t_2) \hat{\boldsymbol{\xi}}{}^{T}(t_1) } \equiv g(t_1) g(t_2) \boldsymbol{\nu}(t_1-t_2)
\end{align}
with the noise kernel \cite{fleming2011b}
\begin{align} \label{eq:nu}
  \boldsymbol{\nu}(t) \equiv \frac{1}{2} \int \limits_{0}^{\infty} \! \mathrm{d} \omega \coth \left( \frac{\beta \omega}{2} \right) \cos ( \omega t ) \mathbf{I}(\omega),
\end{align}
which contains the spectral density \cite{weiss1999}
\begin{align} \label{eq:I}
  \mathbf{I}(\omega) \equiv \mathbf{g}^{T} \mathbf{m}^{-\frac{1}{2}} \omega^{-1} \delta( \omega \mathds{1} - \boldsymbol{\omega} ) \mathbf{m}^{-\frac{1}{2}} \mathbf{g}.
\end{align}
Here we have made use of the Dirac delta distribution $\delta( \omega \mathds{1} - \boldsymbol{\omega} )$ with the identity matrix $\mathds{1}$ and the bath frequency matrix $\boldsymbol{\omega}$, Eq.~{(\ref{eq:omega})}. In brief, the noise kernel describes the noisy influence of the bath on the measurement process. Note that we do not further specify the structure of the spectral density \footnote{A common choice for  Eq.~{(\ref{eq:I})} is an Ohmic spectral density with $\mathbf{I}(\omega) \sim \omega$ and a cut-off term for high frequencies. See, e.\,g., Ref.~\cite{weiss1999} for a more detailed discussion.} and keep it as a general phenomenological expression. Nevertheless, we assume a continuous bath with $N\rightarrow\infty$ for which the definition of a spectral density makes sense in the first place.\par
We furthermore remark that for our chosen initial state $\hat{\varrho}(0)$, Eq.~{(\ref{eq:InitialState})}, the expectation value shift $\mathbf{s}(t)$, Eq.~{(\ref{eq:s})}, vanishes for all times, i.\,e.,
\begin{align} \label{eq:s=0}
  \mathbf{s}(t) = 0.
\end{align}
This means that the first moments of the inferred observables after a certain interaction time $t$ directly correspond to the first moments of the initial system observables, Eq.~{(\ref{eq:inferredexpecation})}. As we have already stated above, this can be considered as a typical behavior.

\subsection{Marginal probability distributions} \label{sec:entropy:marginal probability distributions}
The choice of the initial state $\hat{\varrho}(0)$, Eq.~{(\ref{eq:InitialState})}, allows us to explicitly calculate the the marginal probability distributions, Eq.~{(\ref{eq:Pmarginal})}. It seems natural to assume that these marginal probability distributions do not directly correspond to the initial position distribution $\braket{x|\hat{\varrho}_{\mathrm{S}}(0)|x}$ or the initial momentum distribution $\braket{p|\hat{\varrho}_{\mathrm{S}}(0)|p}$ of the system, but should in addition also incorporate the disturbance from the indirect measurement via the pointers as well as the noisy effects of the bath. Since pointers and bath are initially in a Gaussian state, Eqs.~{(\ref{eq:initialpointerstate})} and {(\ref{eq:initialbathstate})}, their influence on the marginal probability distributions can also be expected to have a Gaussian shape. Indeed, the calculations shown in \ref{sec:appendix:marginal probability distributions} lead us to the broadened distributions
\begin{subequations} \label{eq:Pmarginal2}
  \begin{align} \label{eq:Pmarginal2:X}
    \mathrm{pr}_{\mathcal{X}}(\mathcal{X};t) = \frac{1}{\sqrt{2 \pi} \delta_{\mathcal{X}}(t)} \int \limits_{- \infty}^{+ \infty} \! \mathrm d x \braket{x|\hat{\varrho}_{\mathrm{S}}(0)|x} \exp \left[ - \frac{(\mathcal{X}-x)^2}{2 \delta_{\mathcal{X}}^2(t)} \right]
  \end{align}  
and
  \begin{align} \label{eq:Pmarginal2:P}
    \mathrm{pr}_{\mathcal{P}}(\mathcal{P};t) = \frac{1}{\sqrt{2 \pi} \delta_{\mathcal{P}}(t)} \int \limits_{- \infty}^{+ \infty} \! \mathrm d p \braket{p|\hat{\varrho}_{\mathrm{S}}(0)|p} \exp \left[ - \frac{(\mathcal{P}-p)^2}{2 \delta_{\mathcal{P}}^2(t)} \right],
  \end{align}
\end{subequations}
respectively with the noise terms $\delta_{\mathcal{X}}^2(t)$ and $\delta_{\mathcal{P}}^2(t)$, which represent the diagonal elements of the covariance matrix of the noise operators, Eq.~{(\ref{eq:noiseoperators:cov})}. As also shown in \ref{sec:appendix:marginal probability distributions}, this covariance matrix can be expressed as 
\begin{align} \label{eq:noiseoperators:cov:calc2}
  \begin{pmatrix} \delta_{\mathcal{X}}^2(t) & \delta_{\mathcal{X}\mathcal{P}}(t) \\ \delta_{\mathcal{X}\mathcal{P}}(t) & \delta_{\mathcal{P}}^2(t) \end{pmatrix} = & \phantom{+.} \frac{1}{2} \mathbf{B}(t) \left( \braket{\hat{\mathbf{J}} \hat{\mathbf{J}}^{T}} + \braket{\hat{\mathbf{J}} \hat{\mathbf{J}}^{T}}^{T} \right) \mathbf{B}^{T}(t) \nonumber \\
  & + \int \limits_{0}^{t} \! \mathrm d t_1 \int \limits_{0}^{t} \! \mathrm d t_2 g(t_1) g(t_2) \mathbf{\Lambda}(t,t_1) \boldsymbol{\nu}(t_1-t_2) \mathbf{\Lambda}^{T} (t,t_2).
\end{align}
Here we have recalled the time-dependent coupling strength of the bath $g(t)$, Eq.~{(\ref{eq:g})}, the coefficient matrices $\mathbf{B}(t)$ and $\mathbf{\Lambda}(t,s)$ from the dynamics of the inferred observables, Eq.~{(\ref{eq:inferredsystem})}, the initial value vector $\hat{\mathbf{J}}$, Eq.~{(\ref{eq:J})}, and the noise kernel $\boldsymbol{\nu}(t)$, Eq.~{(\ref{eq:nu})}. Note that the non-diagonal terms in Eq.~{(\ref{eq:noiseoperators:cov:calc2})} correspond to the correlation term $\delta_{\mathcal{X}\mathcal{P}}(t)$, Eq.~{(\ref{eq:noiseoperators:cov})}, which is of no further importance in the following.\par
Briefly summarized, the marginal probability distributions, Eq.~{(\ref{eq:Pmarginal2})}, both consist of a convolution of the system's initial probability distributions with a Gaussian noise function. This noise function itself is determined by a convolution of a Gaussian pointer noise function, Eq.~{(\ref{eq:FphiResult})}, with a Gaussian bath noise function, Eq.~{(\ref{eq:FbathResult})}. In other words, both the pointers and the bath act as a Gaussian filter \cite{buzek1995} through which we are forced to look during the measurement process, and which leave us with a distorted image of the initial probability distributions of the system. The noise terms $\delta_{\mathcal{X}}^2(t)$ and $\delta_{\mathcal{P}}^2(t)$ give a description for this combined disturbance of the measurement results from the pointers and the bath. Since we use Gaussian-shaped initial states for the pointers and the bath, Eqs.~{(\ref{eq:initialpointerstate})} and {(\ref{eq:initialbathstate})}, which are fully characterized by their second moments, the noise terms also contain only second moments: The first term on the right-hand side of Eq.~{(\ref{eq:noiseoperators:cov:calc2})} describes the pointer-based variance contributions to the noise terms, whereas the second term describes the bath-based variance contributions. Both increased pointer-based variance contributions and increased bath-based variance contributions increase the effective disturbance of the marginal probability distributions, Eq.~{(\ref{eq:Pmarginal2})}. In this sense, the noise terms can also be considered as ``extrinsic" or ``measurement uncertainties" in comparison with the ``intrinsic" or ``preparation uncertainties" of the system given by $\braket{x|\hat{\varrho}_{\mathrm{S}}(0)|x}$ and $\braket{p|\hat{\varrho}_{\mathrm{S}}(0)|p}$, respectively. This classification of the disturbance is a concept similarly used for variances as uncertainties of pointer-based simultaneous measurements \cite{appleby1998a,heese2014}.\par
An optimal measurement configuration for a pointer-based measurement is consequently defined by a minimal noise term product. In \ref{sec:appendix:minimal noise term product} we show that the product of the noise terms is bound from below by \footnote{We remark that we have assumed in Ref.~\cite{heese2014} (in Eq.~(55b)) that the pointer-based variance product, i.\,e., the product of the diagonal elements of the first term in Eq.~{(\ref{eq:noiseoperators:cov:calc2})}, is bound from below by $1/4$. While this assumption always holds true for a closed pointer-based measurement due to Ineq.~{(\ref{ineq:dXdP})}, it is in fact not generally valid for an open pointer-based measurement.}
\begin{align} \label{ineq:dXdP}
  \delta_{\mathcal{X}}(t) \delta_{\mathcal{P}}(t) \geq \frac{1}{2},
\end{align}
which defines the best possible accuracy of any measurement apparatus. This statement is closely related to the fact that the variance-based uncertainty of a pointer-based measurement is bound from below by one, see, e.\,g., Ref.~\cite{arthurs1965,arthurs1988,busshardt2010,heese2014}.

\subsection{Lower bound of the collective entropy}
In Ref.~\cite{heese2013} we have discussed the collective entropy for closed pointer-based simultaneous measurements with pure initial system states and, with the help of Ref.~\cite{lieb1978}, have established a lower bound of this collective entropy. Interestingly, although we have used an open-pointer based measurement and possibly mixed initial system states in the present manuscript, the structure of the marginal probability distributions, Eq.~{(\ref{eq:Pmarginal2})}, which determine the collective entropy, Eq.~{(\ref{eq:S})}, is similar to the structure of the marginal probability distribution for closed pointer-based simultaneous measurements. Therefore, with only slight changes, we can adapt the derivation of a lower bound of the collective entropy of closed pointer-based simultaneous measurements to the collective entropy of open pointer-based simultaneous measurements. In the following, we will briefly recapitulate this derivation.\par
First of all, we recall a theorem from Ref.~\cite{lieb1978}, which states that the information entropy
\begin{align} \label{eq:Lieb:S}
  S[ f ] \equiv - \int \limits_{- \infty}^{+ \infty} \! \mathrm d x\ f(x) \ln f(x)
\end{align}
of a Fourier convolution 
\begin{align} \label{eq:ast}
  (f \ast g)(x) \equiv \int \limits_{- \infty}^{+ \infty} \! \mathrm d y f(x-y) g(y)
\end{align}
of two probability distributions $f(x)$ and $g(x)$ is bound from below by
\begin{align} \label{ineq:Lieb:Theorem}
  S[ f * g ] \geq \lambda S[ f ] + ( 1 - \lambda ) S[ g ] - \frac{\lambda \ln \lambda + (1 - \lambda) \ln ( 1 - \lambda )}{2}
\end{align}
with an arbitrary weighting parameter $\lambda \in [0,1]$. Equality in Ineq.~{(\ref{ineq:Lieb:Theorem})} holds true if and only if both $f(x)$ and $g(x)$ are Gaussian distributions with variances $\sigma_f^2$ and $\sigma_g^2$, respectively, and the weighting parameter reads
\begin{align} \label{eqn:GaussianParameter}
  \lambda = \frac{\sigma_f^2}{\sigma_f^2+\sigma_g^2}.
\end{align}
It is clear that the marginal entropies, Eq.~{(\ref{eq:Smarginal})}, which contain the marginal probability distributions, Eq.~{(\ref{eq:Pmarginal2})}, also represent information entropies of convolutions as given by the left-hand side of Ineq.~{(\ref{ineq:Lieb:Theorem})}. Therefore, we can apply the lower bound of Ineq.~{(\ref{ineq:Lieb:Theorem})} to each of the marginal entropies in the collective entropy $S(t)$, Eq.~{(\ref{eq:S})}. In particular, we insert the initial position and momentum distributions of the system in place of $f(x)$ and their associated Gaussian filter functions in place of $g(x)$. For reasons of simplicity, we use the same weighting parameter $\lambda$ for both of these lower bounds. In a next step, we can eliminate the dependency on the system to be measured by making use of the entropic uncertainty relation \cite{everett1956,hirschman1957,bialynicki1975,bialynicki2006}
\begin{align} \label{ineq:Hirsch}
  S[\braket{x|\hat{\varrho}_{\mathrm{S}}(0)|x}] + S[\braket{p|\hat{\varrho}_{\mathrm{S}}(0)|p}] \geq 1 + \ln \pi.
\end{align}
It is based on the Babenko-Beckner inequality \cite{babenko1961,beckner1975}. Saturation in Ineq.~{(\ref{ineq:Hirsch})} occurs only for pure Gaussian states \cite{lieb1990,oezaydin2004}. For a detailed discussion of this and similar entropic uncertainty relations, also see Ref.~\cite{bialynicki2011}. The usage of Ineq.~{(\ref{ineq:Hirsch})}, which contains the density matrix $\hat{\varrho}_{\mathrm{S}}(0)$ of the system to be measured, is the main difference to the derivation from Ref.~\cite{heese2013}, where the system to be measured was limited to pure states and a simplified form of Ineq.~{(\ref{ineq:Hirsch})} with pure states had been used.\par
Accordingly, we arrive at the lower bound
\begin{align} \label{ineq:Lieb:SingleParam}
  S(t) \geq 1 - \lambda \ln \frac{\lambda}{\pi} + (1-\lambda) \ln \left( \frac{2 \pi \delta_{\mathcal{X}}(t) \delta_{\mathcal{P}}(t) }{1-\lambda} \right)
\end{align}
of the collective entropy $S(t)$, Eq.~{(\ref{eq:S})}. In particular, this lower bound depends on the noise terms $\delta_{\mathcal{X}}(t)$ and $\delta_{\mathcal{P}}(t)$, Eq.~{(\ref{eq:noiseoperators:cov:calc2})}. A maximization of the right-hand side of Ineq.~{(\ref{ineq:Lieb:SingleParam})} with respect to $\lambda$ reveals the optimal weighting parameter $\lambda = 1/(1+2\delta_{\mathcal{X}}(t) \delta_{\mathcal{P}}(t))$ and finally leads us to the result
\begin{align} \label{ineq:eur}
  S(t) \geq 1 + \ln \bigg[ 2 \pi \Big( \delta_{\mathcal{X}}(t) \delta_{\mathcal{P}}(t) + \frac{1}{2} \Big) \bigg].
\end{align}
This entropic uncertainty bound for open pointer-based simultaneous measurements is of the same form as the entropic uncertainty bound for closed pointer-based simultaneous measurements from Ref.~\cite{heese2013}. Moreover, it is an extension of the more well-known entropic uncertainty bound $S(t) \geq 1 + \ln (2 \pi)$ from Ref.~\cite{buzek1995}, to which it can be reduced in case of minimal noise terms, Ineq.~{(\ref{ineq:dXdP})}.\par
Due to the equality conditions of Ineqs.~{(\ref{ineq:Lieb:Theorem})} and {(\ref{ineq:Hirsch})}, equality in Ineq.~{(\ref{ineq:eur})} occurs only for initial system states $\hat{\varrho}_{\mathrm{S}}(0)$, which are minimal uncertainty states in the sense of Heisenberg's uncertainty relation \cite{ballentine1998}, i.\,e., pure Gaussian states for which the position variance $\sigma_{x}^2$ and the momentum variance $\sigma_{p}^2$ obey $\sigma_{x}^2 \sigma_{p}^2 = 1/4$; and which additionally fulfill $\sigma_{x}^2 = \delta_{\mathcal{X}}(t)/(2 \delta_{\mathcal{P}}(t))$ for a given set of noise terms \cite{busshardt2011}. Such initial system states can be understood as ``minimal entropy states" \cite{heese2013}. Note that mixed system states with Gaussian density matrices \cite{mann1993} are not sufficient for equality.\par
Summarized, the collective entropy of an open pointer-based measurement, Eq.~{(\ref{eq:S})}, behaves exactly like the collective entropy of a closed pointer-based measurement. Particularly, the collective entropy is bound from below by the sharp entropic uncertainty bound given by Ineq.~{(\ref{ineq:eur})}. The only effect of the environmental heat bath is to modify the collective entropy by modifying the noise terms, Eq.~{(\ref{eq:noiseoperators:cov:calc2})}. An optimal measurement accuracy can be achieved for minimal noise terms, Ineq.~{(\ref{ineq:dXdP})}, and requires a pure Gaussian system state with a specific variance which leads to equality in Ineq.~{(\ref{ineq:eur})}.

%%%%%%%%%%%%%%%%%%%%%%%%%%%%%%%%%%%%%%%%%%%%%%%%%%%%%%%%%%%%%%%%%%%%%%%%%%%%%%%%%%%%%%%%%%%%%%%%%%%%%%%%%%%%%%%%%%%%%%%%%%%%%%%%%%%
%%%%%%%%%%%%%%%%%%%%%%%%%%%%%%%%%%%%%%%%%%%%%%%%%%%%%%%%%%%%%%%%%%%%%%%%%%%%%%%%%%%%%%%%%%%%%%%%%%%%%%%%%%%%%%%%%%%%%%%%%%%%%%%%%%%

\section{Conclusion}
We have shown that it is possible to determine a formal expression for the collective entropy of a general open pointer-based simultaneous measurement. This collective entropy has a very intuitive structure from which the noisy influence of the pointers and the bath on the measurement result can be understood. Moreover, this structure allows us to show that the collective entropy has a sharp lower bound, which can only be reached for specific pure Gaussian system states. In particular, our results are valid for any bilinear interaction between the system and the pointers and any initially mixed system state and thus extend various previous results on this topic.\par
Several simplifications have been made to perform our calculations. First of all, the Hamiltonian only includes bilinear terms. However, we expect that terms of higher order would only lead to relatively small correction terms without fundamentally changing our results. Second, ideal single variable measurements of the pointer observables have to be performed in order to realize the measurement procedure, which is an inherent conceptional weakness of pointer-based simultaneous measurements. Yet we hope that with the help of the environmental heat bath, it might be possible to replace this rather theoretical measurement process by a more natural decoherence process \cite{zurek2003}. Finally, we have used a separable initial state with squeezed states as the initial pointer states and a thermal state as the initial bath state. Although this approach is a clear limitation of our results, we think that our choice is reasonable. It would nevertheless be interesting to discuss different initial states and specifically examine the influence of entanglement and preparation energy on the collective entropy.

%%%%%%%%%%%%%%%%%%%%%%%%%%%%%%%%%%%%%%%%%%%%%%%%%%%%%%%%%%%%%%%%%%%%%%%%%%%%%%%%%%%%%%%%%%%%%%%%%%%%%%%%%%%%%%%%%%%%%%%%%%%%%%%%%%%
%%%%%%%%%%%%%%%%%%%%%%%%%%%%%%%%%%%%%%%%%%%%%%%%%%%%%%%%%%%%%%%%%%%%%%%%%%%%%%%%%%%%%%%%%%%%%%%%%%%%%%%%%%%%%%%%%%%%%%%%%%%%%%%%%%%

\section*{Acknowledgement}
R. H. gratefully acknowledges a grant from the Landes\-graduierten\-f{\"o}rderungs\-gesetz of the state of Baden-W{\"u}rttemberg.

%%%%%%%%%%%%%%%%%%%%%%%%%%%%%%%%%%%%%%%%%%%%%%%%%%%%%%%%%%%%%%%%%%%%%%%%%%%%%%%%%%%%%%%%%%%%%%%%%%%%%%%%%%%%%%%%%%%%%%%%%%%%%%%%%%%
%%%%%%%%%%%%%%%%%%%%%%%%%%%%%%%%%%%%%%%%%%%%%%%%%%%%%%%%%%%%%%%%%%%%%%%%%%%%%%%%%%%%%%%%%%%%%%%%%%%%%%%%%%%%%%%%%%%%%%%%%%%%%%%%%%%

\appendix
\section{Marginal probability distributions} \label{sec:appendix:marginal probability distributions}
In this appendix section we briefly describe how to calculate the marginal probability distributions, Eq.~{(\ref{eq:Pmarginal2})}, from the definition, Eqs.~{(\ref{eq:prXP})} and {(\ref{eq:Pmarginal})}, with the help of the chosen initial state, Eq.~{(\ref{eq:InitialState})}. Our calculations are mainly based on characteristic functions and their Fourier transforms.\par
Inserting the formal dynamics of the inferred observables, Eq.~{(\ref{eq:inferredsystem})}, into the joint probability distribution, Eq.~{(\ref{eq:prXP})}, leads to
\begin{align} \label{eq:prXP0}
  \mathrm{pr}(\mathcal{X},\mathcal{P};t) = \operatorname{tr} \Big\{ & \hat{\varrho}(0) \delta \left[ \mathcal{X} - \hat{X}_{\mathrm{S}}(0) - (1,0) \mathbf{B}(t) \hat{\mathbf{J}} - (1,0) (\mathbf{\Lambda} \star \hat{\boldsymbol{\xi}})(t) \right] \nonumber \\
  & \phantom{..} \times \delta \left[ \mathcal{P} - \hat{P}_{\mathrm{S}}(0) - (0,1) \mathbf{B}(t) \hat{\mathbf{J}} - (0,1) (\mathbf{\Lambda} \star \hat{\boldsymbol{\xi}})(t) \right] \Big\}.
\end{align}
The Dirac delta distributions in Eq.~{(\ref{eq:prXP0})} can be expressed as Fourier transforms of unity, i.\,e., $\delta(x) = \mathcal{F} \left\{ 1 \right\}(x)$. Here we use the notation
\begin{align} \label{eq:FT}
  \mathcal{F} \left\{f\right\}(x) \equiv \frac{1}{2 \pi} \int \limits_{- \infty}^{+ \infty} \! \mathrm d \alpha\ \exp[ \operatorname{i} \alpha x ] f(\alpha)
\end{align}
for the Fourier transform of an arbitrary function $f(x)$, and an analogous notation for Fourier transforms of functions of two variables. Thus, the separability of the initial state $\hat{\varrho}(0)$, Eq.~{(\ref{eq:InitialState})}, allows to rewrite Eq.~{(\ref{eq:prXP0})} as
\begin{align} \label{eq:prXP1}
  \mathrm{pr}(\mathcal{X},\mathcal{P};t) = \mathcal{F}\left\{ F_{\mathrm{S}} F_{\mathrm{P}} F_{\mathrm{B}} \right\}(\mathcal{X},\mathcal{P}) = \left( \mathcal{F}\left\{F_{\mathrm{S}}\right\} \ast \mathcal{F}\left\{F_{\mathrm{P}}\right\} \ast \mathcal{F}\left\{F_{\mathrm{B}}\right\} \right)(\mathcal{X},\mathcal{P})
\end{align}
with the characteristic function of the system
\begin{align} \label{eq:Fpsi}
  F_{\mathrm{S}}(\alpha_1,\alpha_2) \equiv \operatorname{tr} \left\{ \hat{\varrho}_{\mathrm{S}}(0) \exp[ -\operatorname{i} \{ \alpha_1 \hat{X}_{\mathrm{S}}(0) + \alpha_2 \hat{P}_{\mathrm{S}}(0) \} ] \right\},
\end{align}
the characteristic function of the pointers
\begin{align} \label{eq:Fphi}
  F_{\mathrm{P}}(\alpha_1,\alpha_2) \equiv \bra{\sigma_1} \otimes \bra{\sigma_2} \exp[ -\operatorname{i} \{ (\alpha_1,\alpha_2) \mathbf{B}(t) \hat{\mathbf{J}} \} ] \ket{\sigma_1} \otimes \ket{\sigma_2},
\end{align}
and the characteristic function of the bath
\begin{align} \label{eq:Fbath}
  F_{\mathrm{B}}(\alpha_1,\alpha_2) \equiv \operatorname{tr} \left\{ \hat{\varrho}_{\mathrm{B}}(0) \exp[ -\operatorname{i} \{ (\alpha_1,\alpha_2) (\mathbf{\Lambda} \star \hat{\boldsymbol{\xi}})(t) \} ] \right\}.
\end{align}
The symbols $\star$ and $\ast$ are defined in Eqs.~{(\ref{eq:star})} and {(\ref{eq:ast})}, respectively.\par
In the following, we calculate the Fourier transforms of the characteristic functions, Eqs.~{(\ref{eq:Fpsi})} to {(\ref{eq:Fbath})}, one after another with the help of Wigner functions \cite{schleich2001}. This allows us to explicitly perform the convolution in Eq.~{(\ref{eq:prXP1})}. By straightforward integration of the resulting Gaussian expressions we finally arrive at the marginal probability distributions, Eq.~{(\ref{eq:Pmarginal2})}.

\subsection{System}
First of all, the Fourier transform of the characteristic function of the system, Eq.~{(\ref{eq:Fpsi})}, corresponds to the respective Wigner function
\begin{align} \label{eq:FpsiResult}
  W_{\mathrm{S}}(x,p) = \mathcal{F}\left\{F_{\mathrm{S}}\right\}(x,p)
\end{align}
of the system state. We can also use this connection between characteristic function and Wigner function to determine the Fourier transforms of the other two characteristic functions, Eqs.~{(\ref{eq:Fphi})} and {(\ref{eq:Fbath})}, in the following.

\subsection{Pointers}
The Wigner function $W_{\sigma}$ of a squeezed vacuum state $\ket{\sigma}$ with position variance $\sigma^2$ can be written as
\begin{align} \label{eq:Wsigma}
  W_{\sigma}(x,p) \equiv \mathcal{G}\left(-\frac{1}{2 \sigma^2},-2 \sigma^2,0,\frac{1}{\pi};x,p\right),
\end{align}
where we have introduced the general Gaussian function
\begin{align} \label{eq:Gaussian}
  \mathcal{G}(a,b,c,n;x,p) \equiv n \exp \left[ a x^2 + b p^2 + c x p \right].
\end{align}
Equation~{(\ref{eq:Wsigma})} corresponds to the Fourier transform of the generic characteristic function
\begin{align}
  F_{\sigma}'(\alpha_1,\alpha_2) \equiv \braket{ \sigma | \exp[ -\operatorname{i} ( \alpha_1 \hat{x} + \alpha_2 \hat{p} ) ] | \sigma }
\end{align}
of squeezed vacuum states as we have used them for the pointer states, Eq.~{(\ref{eq:initialpointerstate})}, where $\hat{x}$ and $\hat{p}$ stand for the position and momentum observables, respectively, in the Hilbert space of $\ket{\sigma}$. Looking at this relation the other way round yields
\begin{align}
  F_{\sigma}'(x,p) = \mathcal{F}^{-1}\left\{ W_{\sigma} \right\}(x,p) = \mathcal{G}\left(-\frac{\sigma^2}{2},-\frac{1}{8 \sigma^2},0,1;x,p\right),
\end{align}
where $\mathcal{F}^{-1}\left\{f\right\}(x)$ denotes the inverse Fourier transform of an arbitrary function $f(x)$ with $\mathcal{F}\left\{\mathcal{F}^{-1}\left\{f\right\}\right\}(x) = f(x)$, Eq.~{(\ref{eq:FT})}. Thus, the characteristic function of the pointers, Eq.~{(\ref{eq:Fphi})}, can be written as
\begin{align}
  F_{\mathrm{P}}(\alpha_1,\alpha_2) & = F_{\sigma_1}'( (\alpha_1,\alpha_2) \mathbf{B}(t) (1,0,0,0)^{T} , (\alpha_1,\alpha_2) \mathbf{B}(t) (0,0,1,0)^{T} ) \nonumber \\
  & \phantom{=.} \times F_{\sigma_2}'( (\alpha_1,\alpha_2) \mathbf{B}(t) (0,1,0,0)^{T} , (\alpha_1,\alpha_2) \mathbf{B}(t) (0,0,0,1)^{T} )
\end{align}
and one has
\begin{align} \label{eq:FphiResult}
  \mathcal{F}\left\{F_{\mathrm{P}}\right\}(x,p) = \mathcal{G}\left(\frac{a_{\mathrm{P}}(t)}{d_{\mathrm{P}}(t)},\frac{b_{\mathrm{P}}(t)}{d_{\mathrm{P}}(t)},\frac{c_{\mathrm{P}}(t)}{d_{\mathrm{P}}(t)},\frac{1}{2 \pi \sqrt{ d_{\mathrm{P}}(t) }};x,p\right) 
\end{align}
with the coefficients
\begin{subequations}
\begin{align} \label{eq:abcp}
  \begin{pmatrix} -2 b_{\mathrm{P}}(t) & c_{\mathrm{P}}(t) \\ c_{\mathrm{P}}(t) & -2 a_{\mathrm{P}}(t) \end{pmatrix} \equiv \mathbf{B}(t) \mathbf{V} \mathbf{B}^{T}(t)
\end{align}
and
\begin{align}
  d_{\mathrm{P}}(t) \equiv 4 a_{\mathrm{P}}(t) b_{\mathrm{P}}(t) - c_{\mathrm{P}}^2(t),
\end{align}
\end{subequations}
where
\begin{align} \label{eq:JJ}
  \mathbf{V} \equiv \frac{1}{2} \left( \braket{\hat{\mathbf{J}} \hat{\mathbf{J}}^{T}} + \braket{\hat{\mathbf{J}} \hat{\mathbf{J}}^{T}}^{T} \right) = \begin{pmatrix} \sigma_1^2 & 0 & 0 & 0 \\ 0 & \sigma_2^2 & 0 & 0 \\ 0 & 0 & \frac{1}{4 \sigma_1^2} & 0 \\ 0 & 0 & 0 & \frac{1}{4 \sigma_2^2} \end{pmatrix}
\end{align}
denotes the symmetrized pointer covariance matrix, which directly follows from the definition of the initial value vector $\hat{\mathbf{J}}$, Eq.~{(\ref{eq:J})}, and the initial pointer states, Eq.~{(\ref{eq:initialpointerstate})}. In conclusion, Eq.~{(\ref{eq:FphiResult})} can be calculated solely from the coefficient matrix $\mathbf{B}(t)$, Eq.~{(\ref{eq:inferredsystem})}, and the initial pointer position variances $\sigma_1^2$ and $\sigma_2^2$, Eq.~{(\ref{eq:initialpointerstate})}.

\subsection{Bath}
The Wigner function of the thermal bath state $\hat{\varrho}_{\mathrm{B}}(0)$, Eq.~{(\ref{eq:initialbathstate})}, reads
\begin{align}
  W_{\mathrm{B}}(\mathbf{q},\mathbf{k}) & \equiv \operatorname{det} \left[ \tanh \left( \frac{\boldsymbol{\omega} \beta}{2} \right) \pi^{-1} \right] \nonumber \\
  & \phantom{\equiv.} \times \exp \Bigg[ - \mathbf{q}^{T} \mathbf{m}^{\frac{1}{2}} \tanh \left( \frac{\boldsymbol{\omega} \beta}{2} \right) \boldsymbol{\omega} \mathbf{m}^{\frac{1}{2}} \mathbf{q} \nonumber \\
  & \phantom{\equiv \times \exp \Bigg[.} - \mathbf{k}^{T} \mathbf{m}^{-\frac{1}{2}} \tanh \left( \frac{\boldsymbol{\omega} \beta}{2} \right) \boldsymbol{\omega}^{-1} \mathbf{m}^{-\frac{1}{2}} \mathbf{k} \Bigg].
\end{align}
Analogously to the calculation of the characteristic function of the pointers, the generic characteristic bath function
\begin{align}
  F_{\mathrm{B}}'(\boldsymbol{\alpha}_{\mathbf{1}},\boldsymbol{\alpha}_{\mathbf{2}}) \equiv \operatorname{tr} \left\{ \hat{\varrho}_{\mathrm{B}}(0) \exp[ -\operatorname{i} ( \boldsymbol{\alpha}_{\mathbf{1}}^T \hat{\mathbf{q}} + \boldsymbol{\alpha}_{\mathbf{2}}^T \hat{\mathbf{k}} ) ] \right\}
\end{align}
of baths in thermal equilibrium, which is connected to the characteristic bath function, Eq.~{(\ref{eq:Fbath})}, via
\begin{align} \label{eq:Fbath1}
  F_{\mathrm{B}}(\alpha_1,\alpha_2) = F_{\mathrm{B}}' \left( (\mathbf{\Lambda} \star \mathbf{u_q})^T(t) \left( \alpha_1, \alpha_2 \right)^T, (\mathbf{\Lambda} \star \mathbf{u_k})^T(t) \left( \alpha_1, \alpha_2 \right)^T \right),
\end{align}
can be written as
\begin{align}
  F_{\mathrm{B}}'(\mathbf{q},\mathbf{k}) & = \mathcal{F}^{-1}\left\{ W_{\mathrm{B}} \right\}(\mathbf{q},\mathbf{k}) \nonumber \\
  & = \exp \Bigg[ - \frac{1}{4} \mathbf{q}^{T} \mathbf{m}^{-\frac{1}{2}} \coth \left( \frac{\boldsymbol{\omega} \beta}{2} \right) \boldsymbol{\omega}^{-1} \mathbf{m}^{-\frac{1}{2}} \mathbf{q} \nonumber \\
  & \phantom{= \exp \Bigg[} - \frac{1}{4} \mathbf{k}^{T} \mathbf{m}^{\frac{1}{2}}  \coth \left( \frac{\boldsymbol{\omega} \beta}{2} \right) \boldsymbol{\omega} \mathbf{m}^{\frac{1}{2}} \mathbf{k}  \Bigg].
\end{align}
Here we have made use of the abbreviations
\begin{align}
  \mathbf{u_q}(t) \equiv - \mathbf{g}^{T}(t) \mathbf{m}^{-\frac{1}{2}} \cos(\boldsymbol{\omega} t) \mathbf{m}^{\frac{1}{2}}
\end{align}
and
\begin{align}
  \mathbf{u_k}(t) \equiv - \mathbf{g}^{T}(t) \mathbf{m}^{-\frac{1}{2}} \sin(\boldsymbol{\omega} t)  \boldsymbol{\omega}^{-1} \mathbf{m}^{-\frac{1}{2}}.
\end{align}
Explicitly performing the Fourier transform of Eq.~{(\ref{eq:Fbath1})} leads to
\begin{align} \label{eq:FbathResult}
  \mathcal{F}\left\{F_{\mathrm{B}}\right\}(x,p) = \mathcal{G}\left(\frac{a_{\mathrm{B}}(t)}{d_{\mathrm{B}}(t)},\frac{b_{\mathrm{B}}(t)}{d_{\mathrm{B}}(t)},\frac{c_{\mathrm{B}}(t)}{d_{\mathrm{B}}(t)},\frac{1}{2 \pi \sqrt{ d_{\mathrm{B}}(t) }};x,p\right) 
\end{align}
with the coefficients
\begin{subequations}
\begin{align} \label{eq:abcb}
  \begin{pmatrix} - 2 b_{\mathrm{B}}(t) & c_{\mathrm{B}}(t) \\ c_{\mathrm{B}}(t) & - 2 a_{\mathrm{B}}(t) \end{pmatrix} \equiv ( \mathbf{\Lambda} \star \mathbf{v} )(t) ( \mathbf{\Lambda} \star \mathbf{v} )^T(t)
\end{align}
and
\begin{align}
  d_{\mathrm{B}}(t) \equiv 4 a_{\mathrm{B}}(t) b_{\mathrm{B}}(t) - c_{\mathrm{B}}^2(t),
\end{align}
\end{subequations}
where
\begin{align} \label{eq:vv}
  \mathbf{v}(t_1) \mathbf{v}^{T}(t_2) \equiv g(t_1) g(t_2) \boldsymbol{\nu}(t_1-t_2)
\end{align}
with the noise kernel $\boldsymbol{\nu}(t)$, Eq.~{(\ref{eq:nu})}.\par
As a result, Eq.~{(\ref{eq:FbathResult})} is determined by the coefficient matrix $\mathbf{\Lambda}(t,s)$, Eq.~{(\ref{eq:inferredsystem})}, and the noise kernel $\boldsymbol{\nu}(t)$, Eq.~{(\ref{eq:nu})}. Note that for a turned off bath (i.\,e., $\mathbf{g}(t) = 0$ for all $t\geq0$), one has $d_{\mathrm{B}}(t) = 0$ and thus Eq.~{(\ref{eq:FbathResult})} is not well-defined. However, Eq.~{(\ref{eq:FbathResult})} can in this case be understood as a Dirac delta distribution so that the following considerations are still applicable with $a_{\mathrm{B}}(t) = 0$, $b_{\mathrm{B}}(t) = 0$ and $c_{\mathrm{B}}(t) = 0$.

\subsection{Coefficients}
In a next step, we show how the pointer coefficients $a_{\mathrm{P}}(t)$, $b_{\mathrm{P}}(t)$ and $c_{\mathrm{P}}(t)$, Eq.~{(\ref{eq:abcp})}, and the bath coefficients $a_{\mathrm{B}}(t)$, $b_{\mathrm{B}}(t)$, and $c_{\mathrm{B}}(t)$, Eq.~{(\ref{eq:abcb})}, are related to the noise terms $\delta_{\mathcal{X}}^2(t)$ and $\delta_{\mathcal{P}}^2(t)$ and the correlation term $\delta_{\mathcal{X}\mathcal{P}}(t)$, Eq.~{(\ref{eq:noiseoperators:cov})}. For this purpose, we recall the noise operators $\hat{N}_{\mathcal{X}}(t)$ and $\hat{N}_{\mathcal{P}}(t)$, Eq.~{(\ref{eq:noiseoperators})}. According to Eqs.~{(\ref{eq:noiseoperators:exp})} and {(\ref{eq:s=0})}, we can simplify their covariance matrix, Eq.~{(\ref{eq:noiseoperators:cov})}, to
\begin{align} \label{eq:noiseoperators:cov2}
  \begin{pmatrix} \delta_{\mathcal{X}}^2(t) & \delta_{\mathcal{X}\mathcal{P}}(t) \\ \delta_{\mathcal{X}\mathcal{P}}(t) & \delta_{\mathcal{P}}^2(t) \end{pmatrix} = \Braket{ \begin{pmatrix} \hat{N}_{\mathcal{X}}^2(t) & \hat{N}_{\mathcal{X}\mathcal{P}}(t) \\ \hat{N}_{\mathcal{X}\mathcal{P}}(t)  & \hat{N}_{\mathcal{P}}^2(t) \end{pmatrix} }.
\end{align}
Here we have also recalled the symmetrized noise operator $\hat{N}_{\mathcal{X}\mathcal{P}}(t)$, Eq.~{(\ref{eq:noiseoperators:NXP})}. A straightforward calculation using the definition of the noise operators, Eq.~{(\ref{eq:noiseoperators})}, shows that
\begin{align} \label{eq:noiseoperators:cov:calc}
  \begin{pmatrix} \delta_{\mathcal{X}}^2(t) & \delta_{\mathcal{X}\mathcal{P}}(t) \\ \delta_{\mathcal{X}\mathcal{P}}(t) & \delta_{\mathcal{P}}^2(t) \end{pmatrix} = \mathbf{B}(t) \mathbf{V} \mathbf{B}^{T}(t) + ( \mathbf{\Lambda} \star \mathbf{v} )(t) ( \mathbf{\Lambda} \star \mathbf{v} )^T(t)
\end{align}
with the abbreviations introduced in Eqs.~{(\ref{eq:JJ})} and {(\ref{eq:vv})}. Here we have also made use of the symmetry relation \cite{fleming2011b}
\begin{align}
  \braket{ \hat{\boldsymbol{\xi}}(t_1) \hat{\boldsymbol{\xi}}{}^{T}(t_2) } = \braket{ \hat{\boldsymbol{\xi}}(t_1) \hat{\boldsymbol{\xi}}{}^{T}(t_2) }^{T}
\end{align}
for the stochastic force $\hat{\boldsymbol{\xi}}(t)$, Eq.~{(\ref{eq:stochasticforce})}. A comparison of Eq.~{(\ref{eq:noiseoperators:cov:calc})} with Eqs.~{(\ref{eq:abcp})} and {(\ref{eq:abcb})} immediately reveals
\begin{align} \label{eq:noiseterms}
  \begin{pmatrix} \delta_{\mathcal{X}}^2(t) & \delta_{\mathcal{X}\mathcal{P}}(t) \\ \delta_{\mathcal{X}\mathcal{P}}(t) & \delta_{\mathcal{P}}^2(t) \end{pmatrix} = \begin{pmatrix} - 2 ( b_{\mathrm{P}}(t) + b_{\mathrm{B}}(t) ) & c_{\mathrm{P}}(t) + c_{\mathrm{B}}(t) \\ c_{\mathrm{P}}(t) + c_{\mathrm{B}}(t) & - 2 ( a_{\mathrm{P}}(t) + a_{\mathrm{B}}(t) ) \end{pmatrix},
\end{align}
which relates the noise and correlation terms on the left-hand side with the pointer and bath coefficients on the right-hand side.

\subsection{Probability distributions}
At this point, we can collect our previous results. By inserting Eqs.~{(\ref{eq:FpsiResult})}, {(\ref{eq:FphiResult})} and {(\ref{eq:FbathResult})} into Eq.~{(\ref{eq:prXP1})}, we arrive at the final expression for the joint probability distribution
\begin{align} \label{eq:prXP2}
  \mathrm{pr}(\mathcal{X},\mathcal{P};t) = \left( W_{\mathrm{S}} \ast \mathcal{G}\left(-\frac{1}{2 \Delta_{\mathcal{X}}^2(t)},-\frac{1}{2 \Delta_{\mathcal{P}}^2(t)},\gamma(t),\frac{1}{2 \pi \sqrt{d(t)}}\right) \right) (\mathcal{X},\mathcal{P})
\end{align}
with the coefficients
\begin{subequations} \label{eq:coefficients}
\begin{align}
  \Delta_{\mathcal{X}}^2(t) \equiv \frac{-d(t)}{2 ( a_{\mathrm{P}}(t) + a_{\mathrm{B}}(t) ) },
\end{align}
\begin{align}
  \Delta_{\mathcal{P}}^2(t) \equiv \frac{-d(t)}{2 ( b_{\mathrm{P}}(t) + b_{\mathrm{B}}(t) ) },
\end{align}
\begin{align}
  \gamma(t) \equiv \frac{ c_{\mathrm{P}}(t) + c_{\mathrm{B}}(t) }{ d(t) },
\end{align}
and
\begin{align}
  d(t) \equiv 4 ( a_{\mathrm{P}}(t) + a_{\mathrm{B}}(t) ) ( b_{\mathrm{P}}(t) + b_{\mathrm{B}}(t) ) - ( c_{\mathrm{P}}(t) + c_{\mathrm{B}}(t) )^2.
\end{align}
\end{subequations}
Finally, the marginal probability distributions, Eq.~{(\ref{eq:Pmarginal2})}, follow from Eq.~{(\ref{eq:prXP2})} by straightforward integration as defined in Eq.~{(\ref{eq:Pmarginal})} when we make use of the marginals
\begin{subequations}
\begin{align}
  \int \limits_{- \infty}^{+ \infty} \! \mathrm d p W_{\mathrm{S}}(x,p) = \braket{x|\hat{\varrho}_{\mathrm{S}}(0)|x}
\end{align}
and
\begin{align}
  \int \limits_{- \infty}^{+ \infty} \! \mathrm d x W_{\mathrm{S}}(x,p) = \braket{p|\hat{\varrho}_{\mathrm{S}}(0)|p}
\end{align}
\end{subequations}
of the initial Wigner function of the system $W_{\mathrm{S}}(x,p)$, Eq.~{(\ref{eq:FpsiResult})}, where $\braket{x|\hat{\varrho}_{\mathrm{S}}(0)|x}$ and $\braket{p|\hat{\varrho}_{\mathrm{S}}(0)|p}$ stand for the initial position distribution and the initial momentum distribution of the system state, respectively. In particular, the resulting marginal probability distributions, Eq.~{(\ref{eq:Pmarginal2})}, contain the noise terms $\delta_{\mathcal{X}}^2(t)$ and $\delta_{\mathcal{P}}^2(t)$, Eq.~{(\ref{eq:noiseterms})}.

\section{Minimal noise term product} \label{sec:appendix:minimal noise term product}
We show in this appendix section that the product of the noise terms $\delta_{\mathcal{X}}^2(t)$ and $\delta_{\mathcal{X}}^2(t)$, Eq.~{(\ref{eq:noiseoperators:cov})}, obeys Ineq.~{(\ref{ineq:dXdP})}. Our proof is closely related to the considerations in Ref.~\cite{arthurs1988}. Due to the definition of the noise terms, Eq.~{(\ref{eq:noiseoperators:cov})}, we can make use of Robertson's uncertainty relation \cite{robertson1929}, to establish the lower bound
\begin{align} \label{ineq:dxdpRobertson}
  \delta_{\mathcal{X}}(t) \delta_{\mathcal{P}}(t) \geq \frac{1}{2} \left| \Braket{ \left[ \hat{N}_{\mathcal{X}}(t), \hat{N}_{\mathcal{P}}(t) \right] } \right|,
\end{align}
which contains the commutator
\begin{align} \label{eq:Ncommutator}
  \left[ \hat{N}_{\mathcal{X}}(t), \hat{N}_{\mathcal{P}}(t) \right] = \left[ \hat{\mathcal{X}}(t), \hat{\mathcal{P}}(t) \right] - \left[ \hat{\mathcal{X}}(t), \hat{P}_{\mathrm{S}}(0) \right] - \left[ \hat{X}_{\mathrm{S}}(0), \hat{\mathcal{P}}(t) \right]  + \left[ \hat{X}_{\mathrm{S}}(0), \hat{P}_{\mathrm{S}}(0) \right]
\end{align}
of the noise operators $\hat{N}_{\mathcal{X}}(t)$ and $\hat{N}_{\mathcal{P}}(t)$, Eq.~{(\ref{eq:noiseoperators})}. The first commutator on the right-hand side of Eq.~{(\ref{eq:Ncommutator})} vanishes due to Eq.~{(\ref{eq:XPcommute})}. The second and third commutators read
\begin{align}
  \left[ \hat{\mathcal{X}}(t), \hat{P}_{\mathrm{S}}(0) \right] = \left[ \hat{X}_{\mathrm{S}}(0), \hat{\mathcal{P}}(t) \right] = \left[ \hat{X}_{\mathrm{S}}(0), \hat{P}_{\mathrm{S}}(0) \right],
\end{align}
since only the first term on the right-hand side of Eq.~{(\ref{eq:inferredsystem})} acts in the initial system's Hilbert space. As a result, we have
\begin{align} \label{eq:Ncommutator:1}
  \left[ \hat{N}_{\mathcal{X}}(t), \hat{N}_{\mathcal{P}}(t) \right] = - \left[ \hat{X}_{\mathrm{S}}(0), \hat{P}_{\mathrm{S}}(0) \right] = -i.
\end{align}
Inserting Eq.~{(\ref{eq:Ncommutator:1})} into Ineq.~{(\ref{ineq:dxdpRobertson})} finally leads to Ineq.~{(\ref{ineq:dXdP})}.\par
We remark that it would also be possible to use Schr{\"o}dinger's uncertainty relation \cite{schroedinger1930} instead of Robertson's uncertainty relation, Ineq.~{(\ref{ineq:dxdpRobertson})}. This approach would additionally incorporate the correlation term $\delta_{\mathcal{X}\mathcal{P}}(t)$, Eq.~{(\ref{eq:noiseoperators:cov})}, on the right-hand side of Ineq.~{(\ref{ineq:dXdP})}, i.\,e.,
\begin{align}
  \delta_{\mathcal{X}}^2(t) \delta_{\mathcal{P}}^2(t) \geq \delta_{\mathcal{X}\mathcal{P}}^2(t) + \frac{1}{4}.
\end{align}
We do not discuss this extension to Ineq.~{(\ref{ineq:dXdP})} in more detail. It could, however, serve as an interesting point of origin for further considerations.

%%%%%%%%%%%%%%%%%%%%%%%%%%%%%%%%%%%%%%%%%%%%%%%%%%%%%%%%%%%%%%%%%%%%%%%%%%%%%%%%%%%%%%%%%%%%%%%%%%%%%%%%%%%%%%%%%%%%%%%%%%%%%%%%%%%
%%%%%%%%%%%%%%%%%%%%%%%%%%%%%%%%%%%%%%%%%%%%%%%%%%%%%%%%%%%%%%%%%%%%%%%%%%%%%%%%%%%%%%%%%%%%%%%%%%%%%%%%%%%%%%%%%%%%%%%%%%%%%%%%%%%

% REFERENCES
\newpage
\section*{References}
\bibliography{concept}

% DOCUMENT ENDS
\end{document}